 \documentclass[prb,twocolumn]{revtex4}

\usepackage{graphicx}
\usepackage{dcolumn}
\usepackage{bm}
\usepackage[utf8]{inputenc}
\usepackage{color}

\begin{document}

\title{Noncovalent functionalization of carbon nanotubes and graphene with tetraphenylporphyrins: 
Stability and optical properties from {\it ab-initio} calculations}

\author{Walter Orellana\footnote{E-mail: worellana@unab.cl}}
\affiliation{Departamento de Ciencias F\'{i}sicas, Universidad Andres Bello, Rep\'ublica 220, 837-0134 Santiago, Chile}

\author{J. D. Correa}
\affiliation{Departamento de Ciencias B\'asicas, Universidad de Medell\'in, Medell\'in, Colombia}

\date{October 15, 2014}

\begin{abstract}
The stability, electronic and optical properties of single-walled carbon nanotubes
(CNTs) and graphene noncovalently functionalized with free-base tetraphenylporphyrin
(TPP) molecules is addressed by density functional theory calculations, including corrections 
to dispersive interactions. We study the
TPP physisorption on 42 CNT species, particularly those with chiral indices ($n$,$m$),
where $5 \leq n \leq 12$ and $0\leq m\leq n$. Our results show a quite strong $\pi$-$\pi$
interaction between TPP and the CNT surface, with binding energies ranging from 1.1 to
1.8~eV, where higher energies can be associated with increasing CNT diameters. We
also find that the TPP optical absorptions would not be affected by the CNT diameter 
or chirality. Results for the TPP physisorption on graphene show a remarkable stability 
with binding energy of 3.2~eV, inducing a small redshift on the $\pi$-stacked TPP 
absorption bands. The strong graphene-TPP interaction also induces a charge transfer 
from TPP to graphene, indicating a $n$-type doping mechanism without compromising the 
graphene structure.
\end{abstract}

\maketitle

\section{Introduction}

Noncovalent functionalization of single-walled carbon nanotubes (CNTs) with
photoactive molecules is becoming a promising technique to explore functional
materials for light-harvesting or optoelectronic applications.
\cite{Guldi_05,Hecht_06,Mhuir_06,Roquelet_10,Malic_11,Roquelet_11,Gupta_11,Roquelet_12}
The coupling of the exceptional CNT transport properties with the optical
properties of functional dyes, like porphyrins with a strong absorption in
the near infrared and visible, make of CNT/dye complexes good candidates for
sensitive nanoscale devices with potential applications in biomedical imaging,
\cite{Barone_05,Heller_09,Liu_09} and hybrid organic-inorganic photovoltaic devices.
\cite{Campidelli_06}
However, the strength of the $\pi$-stacking interaction with respect to the
CNT structural parameters, like the diameter and chirality, is still lacking.
In addition, a better understanding of the CNT/dye optical properties considering
changes in the dye structure due to the $\pi$-conjugation is needed.

Recently, the kinetics and thermodynamics properties of non-covalently bound CNT/porphyrin
oligomers have been investigated by UV/visible spectroscopy and fluorescence
titration.\cite{Sprafke_11} It was reported that the affinity of the CNTs increases 
sharply with the porphyrin coverage, showing strongest binding energies for semiconducting 
CNTs, particularly those  with chiral indices (7,5) and (8,6). In addition, others works 
have also investigated a supposed selective interaction of porphyrins towards 
semiconducting CNTs, suggesting that the semiconducting and metallic CNTs would have 
significantly different surface properties.\cite{Li_04,Lu_07} These works have also 
suggested that noncovalent functionalization of a mixture of CNTs with porphyrins may 
be an effective method for the separation of semiconducting CNTs from metallic CNTs.
Previously, we reported physisorption properties of iron porphyrin on both metallic
and semiconducting CNTs with similar diameters.\cite{Ruiz_10} Our results showed a
surprising strong $\pi$-stacking interaction, but with negligible energy differences.
Similar results were found for the physisorption strength of free-base and zinc 
porphyrins on the same semiconducting CNT,\cite{Correa_12,Correa_13} showing no clue 
on the suspected selectivity.

Considering the increasing interest in the noncovalent complexation of CNT-based materials
for optoelectronic applications, we carry out an in-depth investigation on the
stability and optical response of free-base tetraphenyporphyrin (TPP) molecules
$\pi$-stacked on single-walled CNTs with different chiralities and diameters,
and also on graphene. Our purpose is to give a theoretically insight into the binding
strength and optical properties of these compounds as a function of the CNT structural
parameters. Our results show that the diameter instead chirality would be the
relevant CNT parameter to explain the unusual strength of the TPP $\pi$-stacking
interaction. In addition, the TPP binding energy on metallic and semiconducting CNTs
does not exhibit important differences that might suggest distinct interactions.
Regarding optical properties, we find that transition bands of the $\pi$-stacked TPP remain
at almost the same energy position than those found in the isolated TPP, being
independent of the CNT structural parameters. For the TPP physisorption on graphene, we
find a very strong binding energy, of about 3.2~eV, providing an upper limit for the
stability of CNT-TPP compounds. We also find a $p$-type doping on graphene induced
by the TPP adsorption.

\section{Theoretical approach}

Our density functional theory (DFT) calculations were performed using the SIESTA 
{\it ab initio} package,\cite{Siesta} which employ norm-conserving pseudopotentials 
and localized atomic orbitals as basis set (double-$\zeta$, singly polarized). The 
physisorption of TPP on both the CNT sidewall and graphene is assessed by van der 
Waals density functional (vdW-DF) as proposed by Dion {\it et al.}\cite{Dion_2004} This 
approach has been successfully applied to describe the dispersive interaction of 
aromatic molecules on the graphite surface, showing good agreement with available 
experimental data.\cite{Chakarova_2006}

In recent experiments, the functionalization of CNTs with TPP molecules has been achieved
by the micelle swelling methods in water suspension,\cite{Roquelet_10}  where the CNT-TPP
compound stays in the micelle core. Because of the hydrophobic character of this core, 
it is not expected water molecules surrounding the compound, which can be considered
close to a vacuum situation. In the present work we study the TPP physisorption 42 
different CNT species, particularly those with chiral indices ($n$,$m$), with 
$5\leq n \leq 12$ and $0\leq m\leq n$. The CNT-TPP compounds are studied within the 
supercell approach, with periodic boundary conditions along the nanotube axis ($z$ 
direction). We use unit cells with a volume of ($30\times30\times\ell a_0$)~\AA$^3$, 
where $a_0$ is CNT lattice constant and $\ell$ a factor to keep a minimum distance 
among TPP images along the CNTs, chosen to be of 10~\AA. This procedure results in supercell 
lengths ($\ell a_0$) between 22-57~\AA, which can contain up to 754 atoms. A grid cutoff 
of 100~Ry and the $\Gamma$ point were used for the real-space and $k$-space integration, 
respectively. For the optical calculations, we use a $1 \times 1 \times 31$ $k$-point 
mesh, for incident light polarized along the CNT axis. The accuracy of these parameters 
were tested considering larger grid cutoff (150~Ry) and $k$-point mesh 
($1\times 1\times 10$). Negligible variations in the total energies, optical spectra, 
and band structures were found, ensuring that our results are converged. For the TPP 
physisorption on graphene, we use a squared unit cell with $6 \times 6$ periodicity and 
the $\Gamma$ point for the $k$-space integration. Here, a vacuum region of 20~\AA\ among
graphene an their images was considered. For the optical response we use a $k$-point
mesh of $31 \times 31 \times 1$, for incident light polarized parallel to the graphene
plane ($x$ and $y$ directions). The TPP binding energy  on the substrates is calculated
by the energy difference between adsorbed and separated constituents, considering
corrections due to the basis set superposition error. The complexes were fully relaxed
by conjugate gradient minimization until the forces on the atoms were less than
0.05~eV/\AA.

The optical response of the functionalized CNTs and graphene is obtained through
first order time dependent perturbation theory, by calculating the imaginary part of
the dielectric function ($\varepsilon_2$). $\varepsilon_2$ gives us a first approach
for the optical absorption coefficient and it is calculated according to the
equation:
{\small\begin{equation}
\varepsilon_2(\omega) = A \int d\mathbf{k}\sum_{c,v} |\hat{\epsilon}\cdot  \langle
\Psi_c({\bf k})|{\bf r}|\Psi_v({\bf k}) \rangle|^2\, \delta (E_c- E_v-\hbar\omega).
\label{e1}
\end{equation}}
Here $A$ is a constant that depends on the cell sizes; $\Psi_c$ and $\Psi_v$ are
the occupied and empty Kohn-Sham orbitals, respectively. The delta function
represents the conservation of energy, which is described by a gaussian function
with a smearing of 0.06~eV.

It is important to note that DFT calculations fail in describing electron-hole
and electron-electron interactions, the so called many-body effects. These effects
are responsible for the formation of excitons and the quasiparticle excitation.
\cite{Mak_11} Therefore, an accurate description of photoexcitations in CNTs and
graphene needs theories beyond DFT, like those based on the GW-Bethe-Salpeter equation
(GW-BSE).\cite{Spataru_04} However, these calculations are not possible right now,
considering the size of the systems under study. Thus, $\varepsilon_2$ can give us a
first approach for the optical response of $\pi$-stacked TPP on CNTs and graphene.
The main difference of DFT calculations with respect to GW-BSE is a redshift in the 
absorption spectrum. For instance, in graphene this redshift is found to be about 
0.6~eV.\cite{Yang_09,Mak_11}

\begin{figure}[t]
\center{\includegraphics[width=6.5cm]{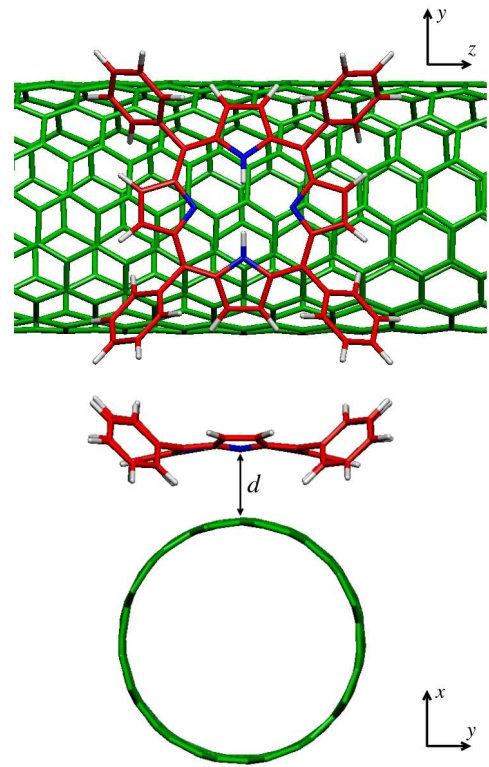}}
\caption{Top and front views of a free-base tetra\-pheny\-lporphyrin
(TPP) physisorbed on the (8,7) CNT in the equilibrium geometry. Blue and white bars
represent nitrogens and hydrogens, respectively, whereas green and red are carbons.}
\label{f1}
\end{figure}

\section{Results and discussion}

\subsection{Energetic and structural properties}

We first discuss different adsorption geometries for the TPP molecule on the CNTs.
Figure~\ref{f1} shows the equilibrium geometry of a TPP adsorbed on a (8,7) CNT. 
Here, the N-H bonds of TPP are oriented perpendicular to the CNT axis, that we 
called position (i). Two other positions for TPP are likely to be found: with the 
N-H bonds parallel to the CNT axis (ii), and with the N-H bonds forming an angle of 
45$^{\circ}$ with respect to the CNT axis (iii). Although position (i) is found to 
be the most stable, the other positions have total energies within 0.3~eV, suggesting 
that all three are equally probables. In the equilibrium geometry, the TPP phenyl 
groups exhibit a small rotation, which come with a molecular twisting. To compare 
the stability of different CNT-TPP complexes, only position (i) will be considered 
throughout this work. The adsorption distance is measured between the CNT surface 
and the N atoms of TPP, as shown in Figure~\ref{f1}. 
\begin{table}[ht]
\caption{Adsorption distance ($d$), binding energy ($E_b$), and bandgap 
energy ($E_g$) of CNT-TPP complexes. $D$ and $\theta_c$ are the CNT diameter
and chiral angle, respectively. Results for TPP on graphene (G-TPP) are also
included.}
\begin{ruledtabular}
\begin{tabular}{llllll}
Complex& $D$(\AA)& $\theta_c$(deg)& $E_g$(eV)& $d$(\AA)& $E_b$(eV) \\
\hline
 (5,0)-TPP  &  3.902  &  0.0  & 0.19  &  3.000  &  1.22 \\
 (5,1)-TPP  &  4.526  &  8.9  & 0.00  &  3.052  &  1.17 \\
 (5,2)-TPP  &  5.083  & 16.1  & 0.00  &  3.106  &  1.18 \\
 (5,3)-TPP  &  5.686  & 21.8  & 1.15  &  3.124  &  1.25 \\
 (5,4)-TPP  &  6.315  & 26.3  & 1.04  &  3.142  &  1.25 \\
  \vspace{0.10cm}
 (5,5)-TPP  &  7.055  & 30.0  & 0.00  &  3.064  &  1.35 \\
 (6,0)-TPP  &  4.921  &  0.0  & 0.00  &  3.117  &  1.11 \\
 (6,1)-TPP  &  5.118  &  7.6  & 0.40  &  3.084  &  1.24 \\
 (6,2)-TPP  &  5.837  & 13.9  & 0.67  &  3.116  &  1.21 \\
 (6,3)-TPP  &  6.401  & 19.1  & 0.02  &  3.180  &  1.21 \\
 (6,4)-TPP  &  7.025  & 23.4  & 1.02  &  3.079  &  1.35 \\
 (6,5)-TPP  &  7.704  & 27.0  & 0.87  &  3.161  &  1.36 \\
  \vspace{0.10cm}
 (6,6)-TPP  &  8.362  & 30.0  & 0.00  &  3.114  &  1.39 \\
 (7,0)-TPP  &  5.685  &  0.0  & 0.24  &  3.091  &  1.18 \\
 (7,1)-TPP  &  6.115  &  6.6  & 0.00  &  3.056  &  1.23 \\
 (7,2)-TPP  &  6.608  & 12.2  & 0.84  &  3.134  &  1.25 \\
 (7,3)-TPP  &  7.165  & 17.0  & 0.86  &  3.183  &  1.24 \\
 (7,4)-TPP  &  7.754  & 21.1  & 0.03  &  3.166  &  1.35 \\
 (7,5)-TPP  &  8.372  & 24.5  & 0.85  &  3.138  &  1.42 \\
 (7,6)-TPP  &  9.038  & 27.5  & 0.75  &  3.142  &  1.50 \\
  \vspace{0.10cm}
 (7,7)-TPP  &  9.703  & 30.0  & 0.00  &  3.182  &  1.48 \\
 (8,0)-TPP  &  6.468  &  0.0  & 0.60  &  3.159  &  1.30 \\
 (8,1)-TPP  &  6.511  &  5.8  & 0.78  &  3.106  &  1.35 \\
 (8,2)-TPP  &  7.384  & 10.9  & 0.09  &  3.143  &  1.37 \\
 (8,3)-TPP  &  7.934  & 15.3  & 0.96  &  3.211  &  1.40 \\
 (8,4)-TPP  &  8.490  & 19.1  & 0.77  &  3.163  &  1.45 \\
 (8,5)-TPP  &  9.103  & 22.4  & 0.02  &  3.192  &  1.46 \\
 (8,6)-TPP  &  9.749  & 25.3  & 0.74  &  3.298  &  1.35 \\
 (8,7)-TPP  & 10.410  & 27.8  & 0.66  &  3.197  &  1.42 \\
  \vspace{0.10cm}
 (8,8)-TPP  & 11.105  & 30.0  & 0.00  &  3.110  &  1.54 \\
 (9,0)-TPP  &  7.269  &  0.0  & 0.15  &  3.213  &  1.33 \\
 (9,1)-TPP  &  7.685  &  5.2  & 1.01  &  3.145  &  1.28 \\
 (9,2)-TPP  &  8.165  &  9.8  & 0.73  &  3.248  &  1.44 \\
  \vspace{0.10cm}
 (9,9)-TPP  & 12.467  & 30.0  & 0.00  &  3.205  &  1.54 \\ 
(10,0)-TPP  &  8.027  &  0.0  & 0.74  &  3.179  &  1.41 \\
(10,1)-TPP  &  8.458  &  4.7  & 0.06  &  3.100  &  1.44 \\
(10,2)-TPP  &  8.938  &  8.9  & 0.85  &  3.126  &  1.44 \\
 \vspace{0.10cm}
(10,10)-TPP & 13.929  & 30.0  & 0.00  &  3.222  &  1.59 \\  
(12,0)-TPP  &  9.638  &  0.0  & 0.10  &  3.180  &  1.46 \\
(12,1)-TPP  & 10.042  &  4.0  & 0.77  &  3.163  &  1.50 \\ 
(12,2)-TPP  & 10.495  &  7.6  & 0.61  &  3.123  &  1.49 \\ 
 \vspace{0.10cm}
(12,12)-TPP & 16.634  & 30.0  & 0.00  &  3.180  &  1.82 \\ 
G-TPP& $\infty$&  -    & 0.00  &  3.424  &  3.32 \\
\end{tabular}
 \end{ruledtabular}
 \label{t1}
 \end{table}
Table~\ref{t1} lists TPP binding energies and binding distances for all the 
complex under study. We also include the DFT bandgaps for the semiconducting 
complexes. Those with small bandgaps ($\sim$0.1~eV) are consistent with empirical 
model predictions of $E_{11}$ as well as with experimental data.\cite{Weisman_03}
We observe that binding energies and adsorption distances vary in ranges given 
by $1.1 \leq E_b \leq 1.8$~eV and $3.0 \leq d \leq 3.3$~\AA, respectively. We 
note that stronger TPP $\pi$-stacking interactions result in larger adsorption 
distance. This can be understood by look at the TPP phenyl groups, whose H atoms
become closer to the CNT surface for large diameter CNTs, as can be seen in
Figure~\ref{f1}. To estimate an upper limit for $E_b$ and $d$, we calculate the TPP 
physisorption on graphene, which can be considered as a CNT with an infinite 
diameter. We find $E_b = 3.32$~eV and $d = 3.42$~\AA. Although the binding distance 
is similar to those found on CNTs with $D \leq 16$~\AA, the binding energy is 
considerably larger, with almost twice the value found on the CNTs. This can be 
understood because of the larger graphene-TPP overlap area, which tend to maximize 
the $\pi$-stacking interaction. 
\begin{figure}[t]
\center{\includegraphics[width=7.0cm]{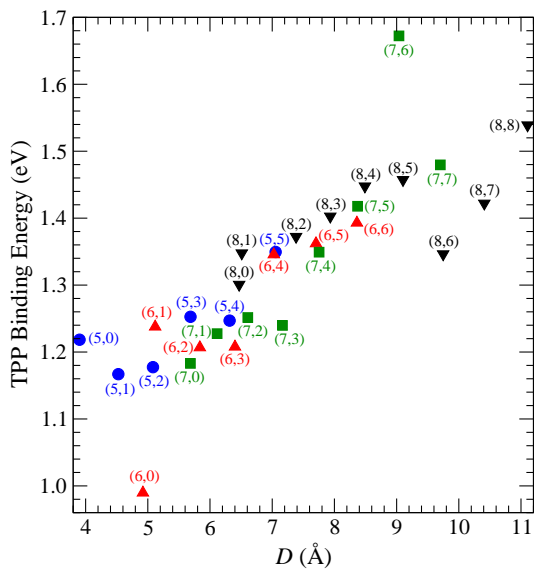}} 
\caption{Binding energy of a TPP molecule physisorbed on different
CNTs as a function of the CNT diameter ($D$).}
\label{f2}
\end{figure}
\begin{figure}[ht]
\center{\includegraphics[width=7.0cm]{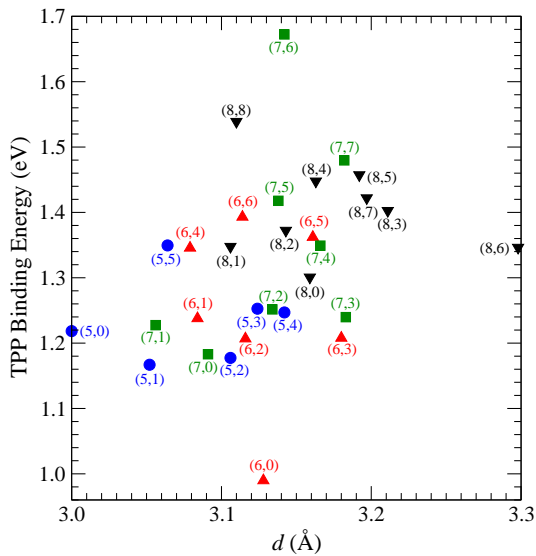}} 
\caption{Binding energy of a TPP molecule physisorbed on different
CNTs as a function of the CNT-TPP binding distance ($d$).}
\label{f3}
\end{figure}
\begin{figure}[ht]
\center{\includegraphics[width=7.0cm]{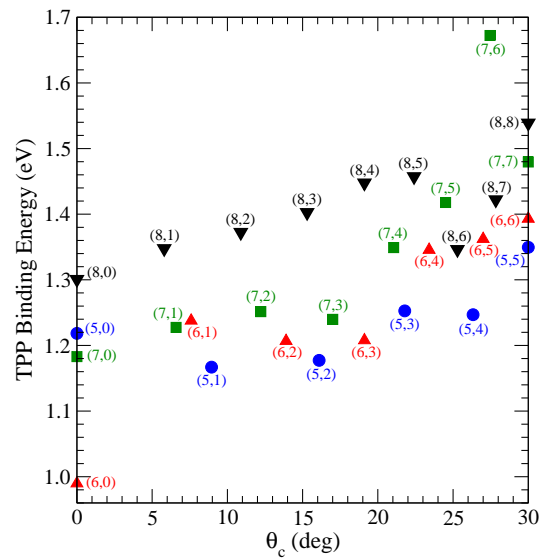}}
\caption{Binding energy of a TPP molecule physisorbed on different
CNTs as a function of the CNT chiral angle ($\theta_c$).}
\label{f4}
\end{figure}
Figure~\ref{f2} shows the CNT-TPP binding energy as a function of the CNT diameter. 
We observe an almost linear increase in the TPP attachment strength with the CNT 
diameter, which in some way confirms our assumption that larger CNT-TPP overlap area 
tends to increase the binding energy of the molecule. However, this general tendency
is not followed in some cases, as can be seen in Fig.~\ref{f2}. For instance, (i) 
the larger TPP binding energy found for the (5,0) CNT ($D=4$~\AA) with respect to 
the (6,0) CNT ($D=5$~\AA), by about 0.2~eV, and (ii) the abrupt changes in the TPP 
binding energy profile going from (6,0) to (6,1), (7,5) to (7,6), and (8,5) to (8,6). 
We attribute these results to the particular atomic geometry of the CNT atoms just 
below the TPP molecule. We observe that the CNT-TPP binding distance tends to 
increase when the H atoms of the phenyl groups nearest neighbors to the CNT surface, 
lie exactly above nanotube C-atoms. This effect is due to the steric repulsion 
between them, which also explain the broad variation on the binding distances between 
the CNT surface and the TPP N-atoms, as shown in Figure~\ref{f3}. 

Figure~\ref{f4} shows the CNT-TPP binding energy as a function of the CNT chiral 
angle. Apparently, higher chiral angles would induce stronger TPP attachment. 
However, if we compare the TPP attachment on CNTs with similar diameters 
and binding distances with those plotted as a function of the chiral angle, we 
observe that they have also similar binding energy, suggesting that the CNT 
diameter instead chirality is the relevant parameter involved in the strong TPP 
binding energy. For instance, if we take the CNTs with chiral indices (5,4) and 
(7,2), they show almost the same TPP binding energy when plotted with respect to 
both the CNT diameter (Fig.~\ref{f2}) and adsorption distance (Fig.~\ref{f3}). 
However, they also show the same binding energy when plotted with respect to 
the chiral angle (Fig.~\ref{f4}). The same behavior can be verified for the 
pairs of CNTs (5,5)-(6,4) and (6,5)-(7,4). These results suggest that the strength
of the TPP interaction on the CNTs would not be related to the CNT surface geometry 
or chirality, but to the CNT diameter.

The adsorption energy of benzene and naphthalene on CNTs (7-8 \AA\ diameter), 
using electron-correlated Moller-Plasset (MP2) method reports binding energies for 
benzene and naphthalene of 0.18 and 0.48 eV, respectively. \cite{Kar_08} This 
results suggest  an overestimation of the TPP adsorption energy on CNTs.
However, temperature programmed desorption experiments performed in ultrahigh 
vacuum have measured the binding energy of polyaromatic hydrocarbons (PAHs) 
adsorbed on graphite, showing quite strong desorption energies. \cite{Zacharia_04}  In 
that work, benzene, naphthalene, coronene, and ovalene adsorbed on graphite show 
desorption energies of 0.50, 0.85, 1.4, and 2.1 eV, respectively. For the case of benzene 
and naphthalene, the experimental values almost double the MP2-theory values. Even 
considering the CNT curvatures, the $\pi$-stacking interaction appears to be much more 
larger than the usually expected. In addition, a recent study of PAHs adsorption on 
graphene shows that the vdW-DF calculations exhibit excellent agreement with the 
available experimental data, while empirical and semiempirical calculations only preserve 
the correct trends.\cite{Bjork_10}   In summary, the work 
concludes that the adsorption  interaction of $\pi$-conjugate systems on graphene are a 
complex combination of dispersive and electrostatic interactions, which would explains 
to some extent our results for the large adsorption energies of TPP on both graphene 
and CNTs.

\subsection{Optical properties}
Noncovalent functionalization of a single-walled (6,5) CNT with TPP molecules 
have been reported in aqueous solution by mean of the micellar suspension
method.\cite{Roquelet2_10} The optical absorption of (6,5)-TPP complexes
shows resonances at 2.357 and 2.831~eV, which are associated to the $Q$ 
and $B$ (Soret) TPP bands, respectively.\cite{Roquelet_10} Our results for
the optical response of the (6,5)-TPP complex for a polarization parallel to the 
tube axis, show clear transitions at 1.60 and 2.04~eV, which we associate to the 
TPP $Q$ and $B$ bands, respectively. The difference in energy between these 
peaks (0.44~eV) is almost the same than those observed in the experiments 
(0.474~eV).\cite{Roquelet_10} However, the theoretical peaks show a redshift of 
about 0.8~eV with respect to the experimental ones, while the shape of the 
absorption spectra would be preserved. This shift is associated to many-body 
effects previously discussed. 
Therefore, through the $\varepsilon_2$ calculation it is possible to investigate, 
qualitatively, changes in the optical properties of different CNT-TPP complexes 
in term of the CNT structural parameters, like diameter or chirality.
\begin{figure}[t]
\center{\includegraphics[width=7.0cm]{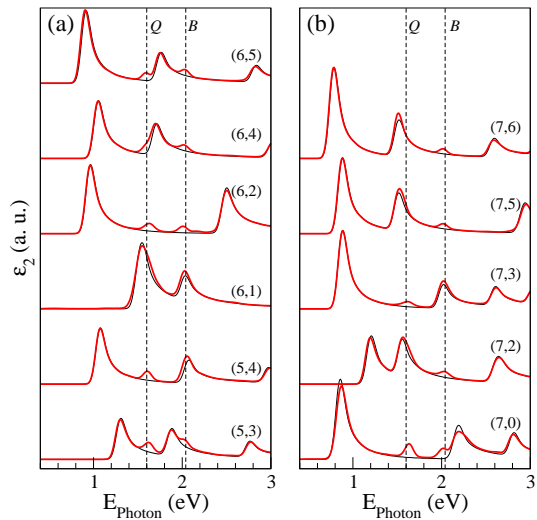}}
\caption{Imaginary part of the dielectric function for semiconducting 
CNTs with a $\pi$-stacked TPP, for incident light polarized parallel to the 
tube axis. Red and black spectra indicate the optical response of CNT-TPP 
and the isolated CNT, respectively. The dashed lines indicate the position of 
$Q$ and $B$ bands of the isolated TPP.}
\label{f5}
\end{figure}

\begin{figure}[t]
\center{\includegraphics[width=7.0cm]{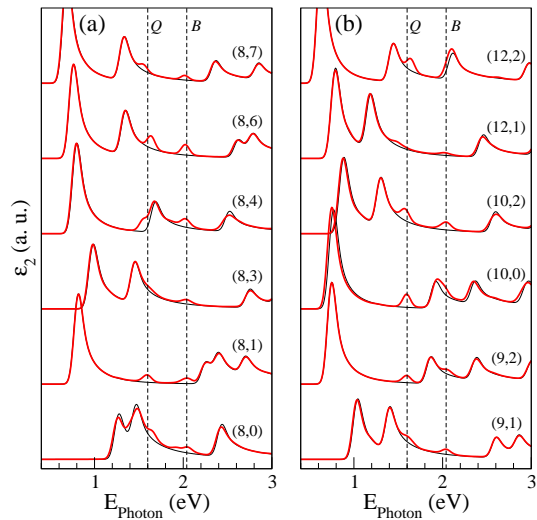}} 
\caption{Imaginary part of the dielectric function for semiconducting 
CNTs with a $\pi$-stacked TPP, for incident light polarized parallel to the 
tube axis. Red and black spectra indicate the optical response of CNT-TPP 
and the isolated CNT, respectively. The dashed lines indicate the position of 
$Q$ and $B$ bands of isolated TPP.}
\label{f6}
\end{figure}

Figure~\ref{f5} and \ref{f6} show $\varepsilon_2$ as a function of the photon 
energy, for incident light polarized parallel to the tube axis, for the
semiconducting CNT-TPP complexes. 
The red and black spectra indicate the optical response of CNT-TPP and the 
isolated CNT, respectively. The vertical dashed lines indicate the position of 
the TPP $Q$ (1.60~eV) and $B$ (2.04~eV) bands. The optical spectra clearly show 
the absorption of the $\pi$-stacked TPP at almost the same position than the 
isolated TPP, indicating that the TPP absorption properties would be preserved in 
the complexes, being independent of the CNT diameter or chirality. However, we must
note that the present results were obtained considering only one TPP molecule
on the CNTs. Increasing TPP coverages may change this conclusion. The optical 
response of complexes considering metallic and semimetallic CNTs (not shown)
exhibit the same properties. Another interesting feature is the resonance between 
the CNT $E_{22}$ transition with the $B$ band for the CNTs: (5,4), (6,1), (7,3), 
and (10,0), and with the $Q$ band for the CNTs: (6,1), (7,5), (7,6), and (7,2).
This tuning between $E_{22}$ and the TPP bands would increase the emission 
of the corresponding complex.

\begin{figure}[ht]
\center{\includegraphics[width=7.0cm]{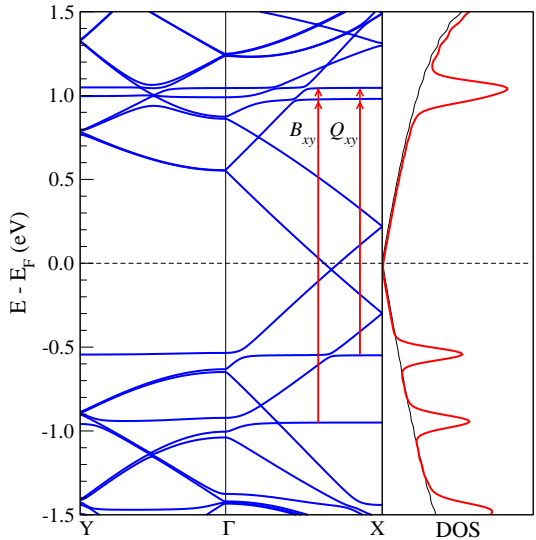}} 
\caption{Band structure and density of states of graphene with 
a $\pi$-stacked TPP represented in a squared reciprocal lattice. The vertical 
arrows indicate the $Q$ and $B$ transitions between $\pi$-stacked TPP frontier 
orbitals. The black DOS corresponds to pristine graphene.}
\label{f7}
\end{figure}

Figure~\ref{f7} shows the electronic band structure and density of states (DOS) 
of the graphene-TPP compound described in a squared reciprocal lattice. The vertical 
arrows show the optical transition between the frontier molecular orbitals of the 
$\pi$-stacked TPP, which correspond to the dispersionless subbands. Here, $Q_x$ 
($Q_y$) is the transition from HOMO to LUMO (LUMO+1), while $B_x$ ($B_y$) is the 
transitions from HOMO-1 to LUMO (LUMO+1). 
These absorptions arise from $\pi-\pi^{*}$ transitions and can be characterized 
approximately by considering only the four molecular levels above mentioned. 
\cite{Gouterman_59,Gouterman_63,Weiss_65,Correa_12} 
In pristine graphene, the valence-band maximum and conduction-band minimum touch 
at one point in the $k$ space, the Dirac point, which occurs at the Fermi energy. 
This indicates that graphene is a zero-gap semiconductor. In Fig.~\ref{f7}, we 
observe that the Fermi energy (the dashed line) is shifted upward with respect to 
the Dirac point energy, which indicates a charge transfer from TPP to graphene, 
revealing a $n$-type doping mechanism. Therefore, graphene would change its 
electronic characteristic induced by the strong TPP complexation. As the 
$\pi$-stacking molecule do not alter the graphene atomic structure, their transport 
properties would be preserved. Similar results have been reported for other 
aromatic molecules physisorbed on graphene.\cite{Mao_13}
\begin{figure}[t]
\center{\includegraphics[width=7.0cm]{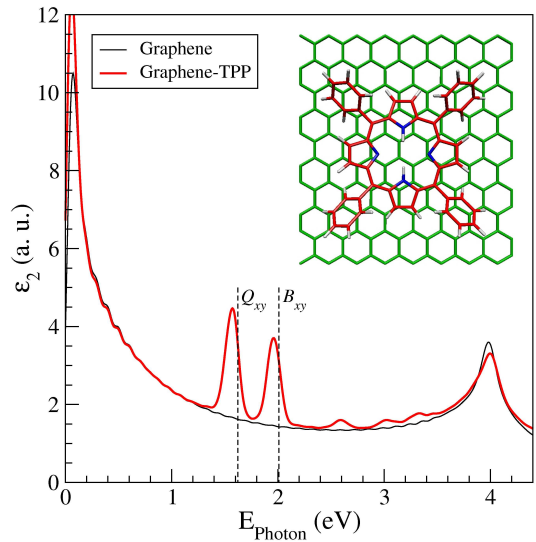}}
\caption{Imaginary part of the dielectric function for graphene 
with a $\pi$-stacked TPP, for in-plain polarization of the incident light. The 
dashed lines indicate the $Q$ and $B$ bands of the isolated TPP.}
\label{f8}
\end{figure}

Figure~\ref{f8} shows $\varepsilon_2$ of a TPP $\pi$-stacked on graphene, The G-TPP 
compound, considering an in-plain polarizations of incident light. For 
pristine graphene, we find a strong absorption in the infrared, below 1~eV, and a 
less intense absorption in the ultraviolet, at around 4~eV, which are attributed to 
transitions among $\pi$ and $\pi^{*}$ bands. These results are in qualitative 
agreement with calculated absorption spectra using the GW-BSE approach as well as 
with experiments,\cite{Yang_09,Chen_11} unless the redshift previously
discussed.
For functionalized graphene, we observe two intense absorptions at around 1.57 and 
1.96~eV, corresponding to the $Q_{xy}$ and $B_{xy}$ bands of the $\pi$-stacked TPP. 
These peaks show a redshift with respect to those of the isolated TPP of 0.05~eV, 
which is associated to the distortion of the TPP phenyl groups due to the strong 
$\pi$-stacking interaction. Additional peaks at around 2.6 and 3.1~eV are also 
associated to the TPP molecule. 

\section{Summary and conclusions}

We have theoretically investigated the stability and optical absorption properties 
of a free-base TPP molecule $\pi$-stacked on single-walled CNTs, as well 
as on graphene. 42 CNTs with chiral indices ($n$,$m$), where $5 \leq n \leq 12$ and 
$0\leq m\leq n$, have been studied. In term of the diameter, they range within
$3.9 \leq D \leq 16.6$~\AA. The TPP physisorption is described by van der Waals 
density functional while the optical properties through the imaginary part of the 
dielectric function in the linear optical response. 

We find a strong CNT-TPP $\pi$-stacking interaction, with binding energies and 
adsorption distance varying within $1.1 \leq E_b \leq 1.8$~eV and 
$3.0 \leq d \leq 3.3$~\AA, respectively, where larger $E_b$ and $d$ are obtained for 
increasing CNT diameters. Our results indicates that diameter instead chirality is 
the relevant parameter for the strong TPP physisorption, which would be originated 
in the increasing overlap area between TPP and the CNTs. An upper limit for the CNT-TPP 
binding strength can be estimated through the TPP physisorption on graphene, which 
can be viewed as an infinite-diameter CNT. We find very strong $\pi$-stacking 
interactions, with $E_b = 3.32$~eV and $d = 3.42$~\AA. However, this strong interaction 
is not enough to open an energy gap in the functionalized graphene, but it can induce a 
charge transfer from TPP, suggesting a $n$-type doping mechanism without compromising
the graphene structure and transport properties. Concerning optical properties, we find 
that $Q$ and $B$ (Soret) TPP absorption bands remain at almost the same energy position 
after the physisorption on the CNTs, being independent of the CNT structural parameters. 
Whereas on graphene, the TPP bands show a redshift of about 0.05~eV, which are 
associated to the distortion of the phenyl groups due to the strong $\pi$-stacking 
interaction. Finally, our results suggest that both compounds CNT-TPP and G-TPP exhibit
remarkable stability while preserving the absorption properties of the chromophore, 
which may be of interest for light-harvesting and bio-labeling applications.

\begin{acknowledgements}
This work was supported by the funding agency CONICYT-PIA under the Grant Anillo~ACT-1107.
JDC acknowledges support from the Universidad de Medell\'in
Research Office through Project No. 684.
\end{acknowledgements} 

\bibliographystyle{spmpsci}

\end{document}